\documentclass[11pt,letterpaper]{article}
\usepackage{emnlp2017}
\usepackage{times}
\usepackage{latexsym}

\usepackage{graphicx} % more modern
\usepackage{subfigure} 

\usepackage{natbib}

\usepackage{algorithm}
\usepackage{algorithmic}

\usepackage{amsmath}
\usepackage{amsfonts}
\usepackage{tabularx}

\emnlpfinalcopy

\def\emnlppaperid{434}

\usepackage{array}
\newcolumntype{H}{>{\setbox0=\hbox\bgroup}c<{\egroup}@{}}

\newcommand{\RR}[0]{\mathbb{R}}

\title{Task-Oriented Query Reformulation with Reinforcement Learning}

\author{Rodrigo Nogueira \\ Tandon School of Engineering\\
New York University\\
		{\tt rodrigonogueira@nyu.edu}
        \And
        Kyunghyun Cho \\ Courant Institute of Mathematical Sciences\\
        Center for Data Science \\
        New York University\\
  {\tt kyunghyun.cho@nyu.edu}}

\date{}

\begin{document} 

\maketitle

\begin{abstract} 

Search engines play an important role in our everyday lives by assisting us in finding the information we need. When we input a complex query, however, results are often far from satisfactory. In this work, we introduce a query reformulation system based on a neural network that rewrites a query to maximize the number of relevant documents returned.
We train this neural network with reinforcement learning. The actions correspond to selecting terms to build a reformulated query, and the reward is the document recall. We evaluate our approach on three datasets against strong baselines and show a relative improvement of 5-20\% in terms of recall. Furthermore, we present a simple method to estimate a conservative upper-bound performance of a model in a particular environment and verify that there is still large room for improvements.
\end{abstract}

\section{Introduction}

Search engines help us find what we need among the vast array of available data. When we request some information using a long or inexact description of it, these systems, however, often fail to deliver relevant items. In this case, what typically follows is an iterative process in which we try to express our need differently in the hope that the system will return what we want. This is a major issue in information retrieval. For instance,~\citet{huang2009analyzing} estimate that 28-52\% of all the web queries are modifications of previous ones.

\begin{figure}[ht]
%\vskip 0.2in
\begin{center}
\centerline{\includegraphics[width=\columnwidth]{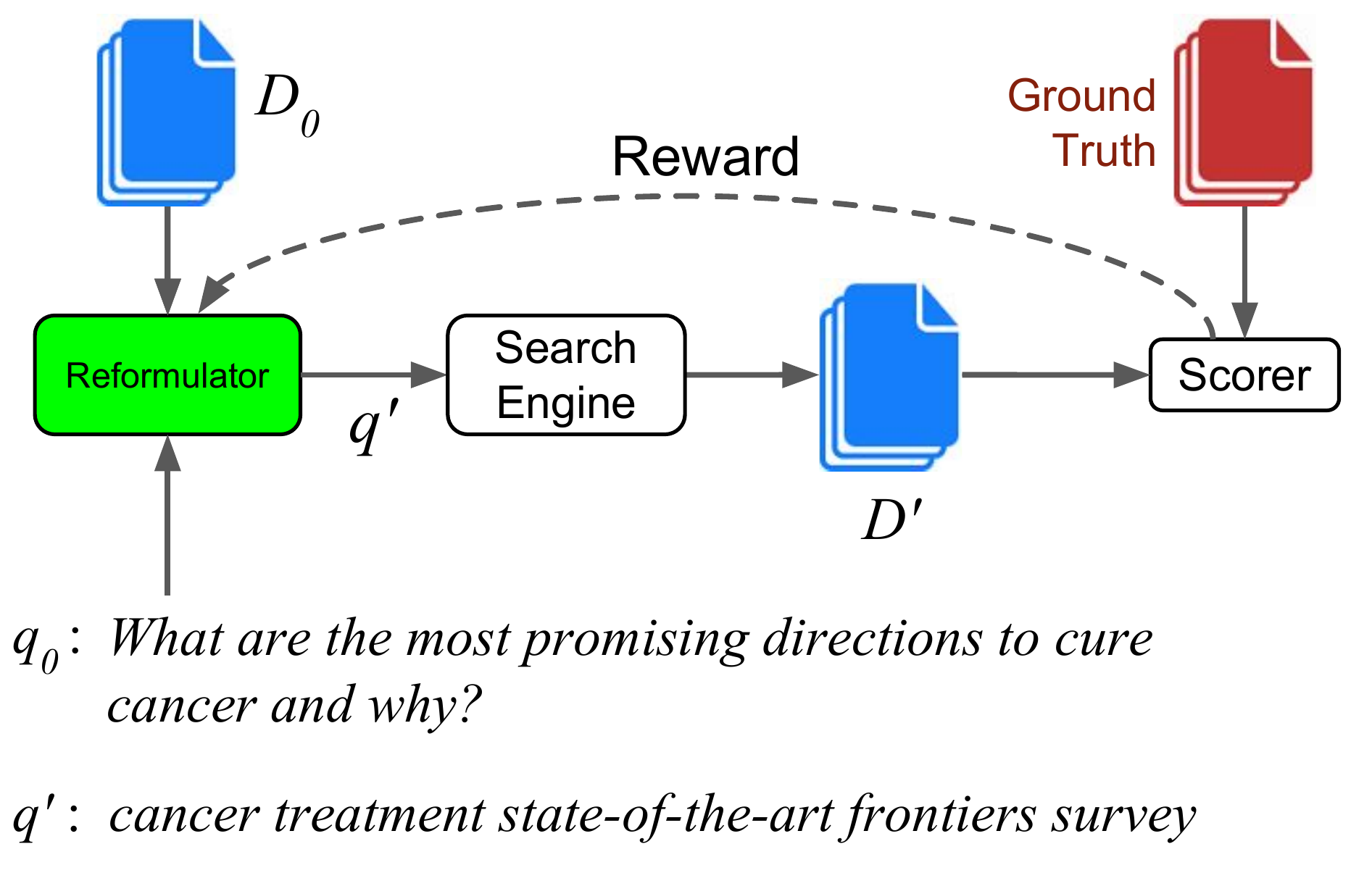}}
%\vskip -0.1in
\caption{A graphical illustration of the proposed framework for query reformulation. A set of documents $D_0$ is retrieved from a search engine using the initial query $q_0$. Our reformulator selects terms from $q_0$ and $D_0$ to produce a reformulated query $q'$ which is then sent to the search engine. Documents $D'$ are returned, and a reward is computed against the set of ground-truth documents.
The reformulator is trained with reinforcement learning to produce a query, or a series of queries, to maximize the expected return.% This process of search and reformulation can be repeated multiple times.
}
\label{fig:query_reformulator}
\end{center}

\vskip -4mm
\end{figure}

To a certain extent, this problem occurs because search engines rely on matching words in the query with words in relevant documents, to perform retrieval. If there is a mismatch between them, a relevant document may be missed.

%According to~\citet{furnas1987vocabulary}, two people use different terms to describe the same concept in more than 80\% of the time.

One way to address this problem is to automatically rewrite a query so that it becomes more likely to retrieve relevant documents. This technique is known as \textit{automatic query reformulation}. It typically expands the original query by adding terms from, for instance, dictionaries of synonyms such as WordNet~\cite{miller1995wordnet}, or from the initial set of retrieved documents~\cite{xu1996query}. This latter type of reformulation is known as pseudo (or blind) relevance feedback (PRF), in which the relevance of each term of the retrieved documents is automatically inferred.

The proposed method is built on top of PRF but differs from previous works as we frame the query reformulation problem as a reinforcement learning (RL) problem. An initial query is the natural language expression of the desired goal, and an agent (i.e. reformulator) \textit{learns} to reformulate an initial query to maximize the expected return (i.e. retrieval performance) through actions (i.e. selecting terms for a new query). The environment is a search engine which produces a new state (i.e. retrieved documents). Our framework is illustrated in Fig.~\ref{fig:query_reformulator}.

The most important implication of this framework is that a search engine is treated as a \textit{black box} that an agent learns to use in order to retrieve more relevant items. This opens the possibility of training an agent to use a search engine for a task other than the one it was originally intended for. To support this claim, we evaluate our agent on the task of question answering (Q\&A), citation recommendation, and passage/snippet retrieval.

As for training data, we use two publicly available datasets (TREC-CAR and Jeopardy) and introduce a new one (MS Academic) with hundreds of thousands of \textit{query}/\textit{relevant document} pairs from the academic domain.

Furthermore, we present a method to estimate the upper bound performance of our RL-based model. Based on the estimated upper bound, we claim that this framework has a strong potential for future improvements.

Here we summarize our main contributions:
\begin{itemize}
\setlength\itemsep{1pt}
\item A reinforcement learning framework for automatic query reformulation.
\item A simple method to estimate the upper-bound performance of an RL-based model in a given environment.
\item A new large dataset with hundreds of thousands of \textit{query}/\textit{relevant document} pairs.\footnote{The dataset and code to run the experiments are available at \url{https://github.com/nyu-dl/QueryReformulator}.}
\end{itemize}

\begin{figure}
%\vskip 0.2in
\begin{center}
\centerline{\includegraphics[width=\columnwidth]{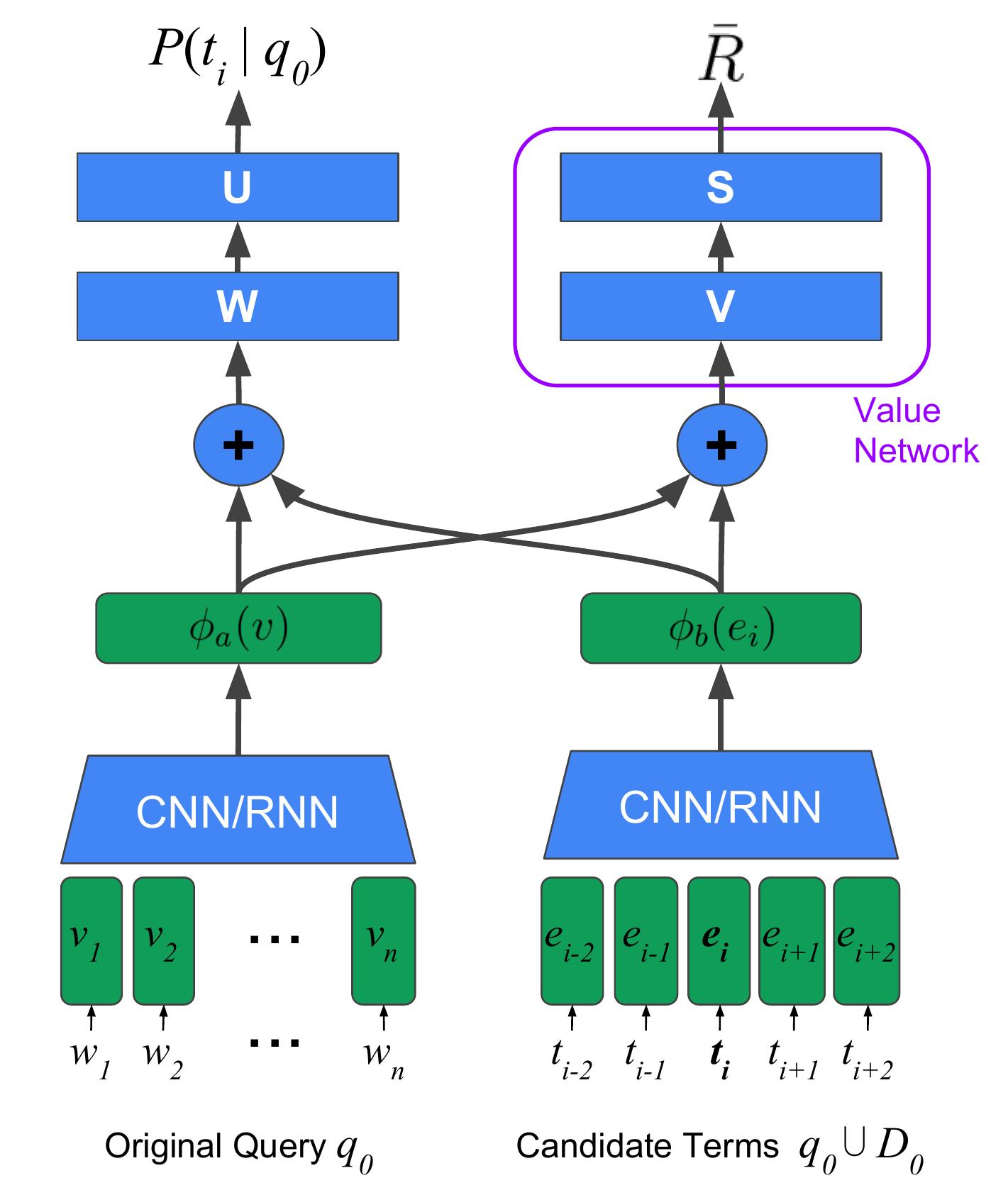}}
\caption{An illustration of our neural network-based reformulator.}
\label{fig:reformulator}
\end{center}

\vskip -4mm
\end{figure} 

\section{A Reinforcement Learning Approach}

\subsection{Model Description}
\label{sec:rl}

In this section we describe the proposed method, illustrated in Fig.~\ref{fig:reformulator}.

The inputs are a query $q_0$ consisting of a sequence of words
$(w_1, ..., w_n)$ and a candidate term $t_i$ with some context words $(t_{i-k},\allowbreak ...,\allowbreak t_{i+k})$, where $k \geq 0$ is the context window size. Candidate terms are from $q_0 \cup D_0$, the union of the terms in the original query and those from the documents $D_0$ retrieved using $q_0$.

We use a dictionary of pretrained word embeddings~\cite{mikolov2013efficient} to convert the symbolic terms ${w_j}$ and ${t_i}$ to their vector representations $v_j$ and $e_i \in \RR^d$, respectively. We map out-of-vocabulary terms to an additional vector that is learned during training.

%Begin: modified after the submission on apr/17
We convert the sequence $\{v_j\}$ to a fixed-size vector $\phi_a(v)$ by using either a Convolutional Neural Network (CNN) followed by a max pooling operation over the entire sequence~\cite{kim2014convolutional} or by using the last hidden state of a Recurrent Neural Network (RNN).\footnote{To deal with variable-length inputs in a mini-batch, we pad smaller ones with zeros on both ends so they end up as long as the largest sample in the mini-batch.}

Similarly, we fed the candidate term vectors ${e_i}$ to a CNN or RNN to obtain a vector representation ${\phi_b(e_i)}$ for each term $t_i$. The convolutional/recurrent layers serve an important role of capturing context information, especially for out-of-vocabulary and rare terms. CNNs can process candidate terms in parallel, and, therefore, are faster for our application than RNNs. RNNs, on the other hand, can encode longer contexts.
%End: modified after the submission on apr/17

Finally, we compute the probability of selecting $t_i$ as:
\begin{equation} \label{eq:1}
P(t_i|q_0) = \sigma( U^\mathsf{T} \tanh( W(\phi_a(v) \Vert \phi_b(e_i)) + b )),
\end{equation}
where $\sigma$ is the sigmoid function, $\Vert$ is the vector concatenation operation, $W \in \RR^{d \times 2d}$ and $U \in \RR^{d}$ are weights, and $b \in \RR$ is a bias.

At test time, we define the set of terms used in the reformulated query as $T=\{t_i\ |\ P(t_i|q_0)>\epsilon\}$, where $\epsilon$ is a hyper-parameter.
At training time, we sample the terms according to their probability distribution, $T=\{t_i\ |\ \alpha=1 \wedge \alpha \sim P(t_i|q_0)\}$.
We concatenate the terms in $T$ to form a reformulated query $q'$, which will then be used to retrieve a new set of documents $D'$. 

%\subsection{Synonyms}

%In addition to the possibility of selecting terms from the feedback documents to
%reformulate the query, the model can also choose terms that are synonyms to
%original query terms. These synonyms are drawn from a lexical database such as
%WordNet~\cite{miller1995wordnet}.

%Begin: added after the submission on apr/17
\subsection{Sequence Generation}
\label{sec:seqgen}
One problem with the method previously described is that terms are selected independently. This may result in a reformulated query that contains duplicated terms since the same term can appear multiple times in the feedback documents. Another problem is that the reformulated query can be very long, resulting in a slow retrieval.

To solve these problems, we extend the model to sequentially generate a reformulated query, as proposed by~\citet{buck2017ask}. We use a Recurrent Neural Network (RNN) that selects one term at a time from the pool of candidate terms and stops when a special token is selected. The advantage of this approach is that the model can remember the terms previously selected through its hidden state. It can, therefore, produce more concise queries.% as we will show in the results section.

We define the probability of selecting $t_i$ as the k-th term of a reformulated query as:
\begin{equation}
P(t_i^k|q_0) \propto \exp(\phi_b(e_i)^\mathsf{T} h_k),
\end{equation}
where $h_k$ is the hidden state vector at the k-th step, computed as:
\begin{equation}
h_k = \tanh(W_a \phi_a(v) + W_b \phi_b(t^{k-1}) + W_h h_{k-1}),
\end{equation}
where $t^{k-1}$ is the term selected in the previous step and $W_a \in \RR^{d \times d}$, $W_b \in \RR^{d \times d}$, and $W_h \in \RR^{d \times d}$ are weight matrices. In practice, we use an LSTM~\cite{hochreiter1997long} to encode the hidden state as this variant is known to perform better than a vanilla RNN.

We avoid normalizing over a large vocabulary by using only terms from the retrieved documents. This makes inference faster and training practical since learning to select words from the whole vocabulary might be too slow with reinforcement learning, although we leave this experiment for a future work.

%End: added after the submission on apr/17

\subsection{Training}

We train the proposed model using REINFORCE~\cite{williams1992simple} algorithm. The per-example stochastic objective is defined as
\begin{equation} \label{eq:2}
    C_a = (R - \bar{R}) \sum_{t \in T} -\log P(t | q_0),
\end{equation}
where $R$ is the reward and $\bar{R}$ is the baseline,
computed by the value network as:
\begin{equation} \label{eq:3}
	\bar{R} = \sigma( S^\mathsf{T} \tanh(V ( \phi_a(v) \Vert \bar{e} ) + b)),
\end{equation}
where $\bar{e} = \frac{1}{N} \sum_{i=1}^{N} \phi_b(e_i)$, $N=|q_0 \cup D_0|$, $V \in \RR^{d \times 2d}$ and $S \in \RR^{d}$ are weights and $b \in \RR$ is a bias.
We train the value network to minimize
\begin{equation} \label{eq:4}
	C_b = \alpha||R-\bar{R}||^2,
\end{equation}
where $\alpha$ is a small constant (e.g. 0.1) multiplied to the loss in order to stabilize learning. We conjecture that the stability is due to the slowly evolving value network which directly affects the learning of the policy. This effectively prevents the value network to fit extreme cases (unexpectedly high or low reward.)

We minimize $C_a$ and $C_b$ using stochastic gradient descent (SGD) with the gradient computed by backpropagation~\cite{rumelhart1988learning}. This allows the entire model to be trained end-to-end directly to optimize the retrieval performance. 

%\paragraph{Value Network Learning}
%To stabilize learning, the gradient for the value network is rescaled by a small constant $\alpha$ (e.g. 0.1). This is equivalent to changing Eq.~\eqref{eq:4} to $C_b = \alpha||R-\bar{R}||^2$. 
%We found in preliminary experiments that this technique has a similar efficacy of the commonly used trick of accumulating gradients over multiple time steps before applying them~\cite{mnih2016asynchronous}. It is, however, simpler to implement since there is no need to keep two separate network weights.

\paragraph{Entropy Regularization}
We observed that the probability distribution in Eq.\eqref{eq:1} became highly peaked in preliminary experiments. This phenomenon led to the trained model not being able to explore new terms that could lead to a better-reformulated query. We address this issue by regularizing the negative entropy of the probability distribution. We add the following regularization term to the original cost function in Eq.~\eqref{eq:2}:
\begin{equation}
	C_H = -\lambda \sum_{t \in q_0 \cup D_0} P(t|q_0)\log P(t|q_0),
\end{equation}
where $\lambda$ is a regularization coefficient.

\section{Related Work}

Query reformulation techniques are either based on a global method, which ignores a set of documents returned by the original query, or a local method, which adjusts a query relative to the documents that initially appear to match the query. In this work, we focus on local methods.
%, although a short empirical analysis of global methods is given in section~\ref{sec:syns}.

A popular instantiation of a local method is the \textit{relevance model}, which incorporates pseudo-relevance feedback into a language model form~\cite{lavrenko2001relevance}. The probability of adding a term to an expanded query is proportional to its probability of being generated by the language models obtained from the original query and the document the term occurs in. This framework has the advantage of not requiring \textit{query/relevant documents} pairs as training data since inference is based on word co-occurrence statistics.

%We argue, however, that for tasks in which the relation of query and document terms go beyond the similarity, such as Q\&A and citation recommendation, models that can learn through examples have a distinct advantage over models that explicitly explore the term similarity relationship, such as relevance model.

%We argue, however, that for tasks that it is not clear how to explicitly model the relevance of a document term to a query, methods that can learn their reformulation strategies from training examples have a distinct advantage.\todo{rewrite this last sentence}

%Our proposal partly lies in the category of Pseudo (or Blind) Relevance Feedback, in which the relevance of the returned documents is automatically inferred \cite{salton1997improving} and partly belongs to the global analysis methods, in which the reformulated query terms are independent of the query and results returned from it, so that changes in the query wording will cause the new query to match other semantically similar terms~\cite{xu1996query}.

Unlike the relevance model, algorithms can be trained with supervised learning, as proposed by~\citet{cao2008selecting}. A training dataset is automatically created by labeling each candidate term as relevant or not based on their individual contribution to the retrieval performance. Then a binary classifier is trained to select expansion terms. In Section~\ref{sec:experiments}, we present a neural network-based implementation of this supervised approach.

%Begin: added after the submission on apr/17
A generalization of this supervised framework is to \textit{iteratively} reformulate the query by selecting one candidate term at each retrieval step. This can be viewed as navigating a graph where the nodes represent queries and associated retrieved results and edges exist between nodes whose queries are simple reformulations of each other~\cite{diaz2016pseudo}. However, it can be slow to reformulate a query this way as the search engine must be queried for each newly added term. In our method, on the contrary, the search engine is queried with multiple new terms at once.  %Moreover, this sequential query reformulation does not account for exploration, that is, a sequential selection of terms that would only return relevant documents after multiple iterations.\todo{REWRITE}
%End: added after the submission on apr/17

An alternative technique based on supervised learning is to learn a common latent representation of queries and relevant documents terms by using a \textit{click-through} dataset~\cite{sordoni2014learning}. Neighboring document terms of a query in the latent space are selected to form an expanded query. Instead of using a \textit{click-through} dataset, which is often proprietary, it is possible to use an alternative dataset consisting of anchor text/title pairs. In contrast, our approach does not require a dataset of paired queries as it learns term selection strategies via reinforcement learning.

Perhaps the closest work to ours is that by \citet{narasimhan2016improving}, in which a reinforcement learning based approach is used to reformulate queries iteratively.
A key difference is that in their work the reformulation component uses domain-specific template queries. Our method, on the other hand, assumes open-domain queries. 
%The task consists of extracting values of a particular event using external information sources to resolve ambiguities. 
%The experiments are carried out on two relatively small datasets of shooting incidents and food adulteration cases.
%A key difference to our approach is that in their work the reformulation component uses domain-specific template queries. Our method, on the other hand, assumes open-domain queries. 

%Begin: added after the submission on apr/17
%Missing information in databases can be automatically populated by a mechanism that uses the web as an external resource and queries that are generated using reinforcement learning~\cite{kanani2012selecting}. This method, however, only addresses information needs that are present in common database schemas, such as \textit{names}, \textit{emails}, and \textit{job titles}. It does not account for more complex needs expressed in natural language as we do.\todo{rewrite}
%End: added after the submission on apr/17

\section{Experiments}
\label{sec:experiments}
In this section we describe our experimental setup, including baselines against which we compare the proposed method, metrics, reward for RL-based models, datasets and implementation details.

\subsection{Baseline Methods}

\paragraph{Raw:} The original query is given to a search engine without any modification. We evaluate two search engines in their default configuration: Lucene\footnote{https://lucene.apache.org/} (Raw-Lucene) and Google Search\footnote{https://cse.google.com/cse/} (Raw-Google). 

\paragraph{Pseudo Relevance Feedback (PRF-TFIDF):}
A query is expanded with terms from the documents retrieved by a search engine using the original query. In this work, the top-$N$ TF-IDF terms from each of the top-$K$ retrieved documents are added to the original query, where $N$ and $K$ are selected by a grid search on the validation data.

\paragraph{PRF-Relevance Model (PRF-RM):} This is a popular relevance model for query expansion by~\citet{lavrenko2001relevance}. The probability of using a term $t$ in an expanded query is given by:
\begin{multline}
P(t|q_0) = (1-\lambda) P'(t|q_0)\\
+ \lambda \sum_{d \in D_0} P(d) P(t|d) P(q_0|d),
\end{multline}
where $P(d)$ is the probability of retrieving the document $d$, assumed uniform over the set, $P(t|d)$ and $P(q_0|d)$ are the probabilities assigned by the language model obtained from $d$ to $t$ and $q_0$, respectively. $P'(t|q_0)= \frac{\text{tf}(t \in q)}{|q|}$, where $\text{tf}(t,d)$ is the term frequency of $t$ in $d$. We set the interpolation parameter $\lambda$ to 0.5, following~\citet{zhai2001study}.

We use a Dirichlet smoothed language model~\cite{zhai2001study} to compute a language model from a document $d \in D_0$:
\begin{equation}
P(t|d)=\frac{\text{tf}(t,d)+u P(t|C)}{|d| + u},
\end{equation}
where $u$ is a scalar constant ($u=1500$ in our experiments), and $P(t|C)$ is the probability of $t$ occurring in the entire corpus $C$.

We use the $N$ terms with the highest $P(t|q_0)$ in an expanded query, where $N$ is a hyper-parameter.

\paragraph{Embeddings Similarity:}
Inspired by the methods proposed by~\citet{roy2016using} and~\citet{,kuzi2016query}, the top-$N$ terms are selected based on the cosine similarity of their embeddings against the original query embedding. 
Candidate terms come from documents retrieved using the original query (PRF-Emb), or from a fixed vocabulary (Vocab-Emb). We use pretrained embeddings from~\citet{mikolov2013efficient}, and it contains 374,000 words.

%In the Rocchio algorithm~\cite{rocchio1971relevance}, the reformulated query is a reweighting of original and retrieved terms that follows the equation:
%\[
%    Q_m = \alpha Q_0 + \beta \frac{1}{|D_r|} \sum_{D_j \in D_r} D_j
%\]
%Where $Q_0$ and $Q_m$ are the one-hot vector representation of the original and reformulated queries, $\alpha$ and $\beta$ are scalars hyper-parameters, and $D_j$ is the one-hot vector representation of the terms in one of the feedback documents.

\subsection{Proposed Methods}
\label{sec:proposed_methods}

\paragraph{Supervised Learning (SL):} Here we detail a deep learning-based variant of the method proposed by~\citet{cao2008selecting}. It assumes that query terms contribute independently to the retrieval performance. We thus train a binary classifier to select a term if the retrieval performance increases beyond a preset threshold when that term is added to the original query. More specifically, we mark a term as relevant if $(R' - R) / R > 0.005$, where $R$ and $R'$ are the retrieval performances of the original query and the query expanded with the term, respectively.

We experiment with two variants of this method: one in which we use a convolutional network for both original query and candidate terms (SL-CNN), and the other in which we replace the convolutional network with a single hidden layer feed-forward neural network (SL-FF). In this variant, we average the output vectors of the neural network to obtain a fixed size representation of $q_0$.

%Begin: modified after the submission on apr/17
\paragraph{Reinforcement Learning (RL):} We use multiple variants of the proposed RL method. RL-CNN and RL-RNN are the models described in Section~\ref{sec:rl}, in which the former uses CNNs to encode query and term features and the latter uses RNNs (more specifically, bidirectional LSTMs). RL-FF is the model in which term and query vectors are encoded by single hidden layer feed-forward neural networks. In the RL-RNN-SEQ model, we add the sequential generator described in Section~\ref{sec:seqgen} to the RL-RNN variant.

%\paragraph{Reinforcement Learning (RL):} We use multiple variants of the proposed RL method: RL-CNN, which is the method described in Section~\ref{sec:rl}, RL-RNN, in which we replace the CNNs by RNNs (more specifically, bidirectional LSTMs), RL-FF, in which we replace the convolutional networks with single hidden layer feed-forward neural networks, and RL-RNN-SEQ, in which the sequential generator described in Section~\ref{sec:seqgen} is added to the RL-RNN variant.\todo{REWRITE}
%In the RL-FF variant, the only source for local context information is from the pretrained embeddings.
%End: modified after the submission on apr/17

\begin{table*}

%\vskip 0.05in
\begin{center}
\begin{small}
\begin{tabular}{lcc|ccc|cc|cc}
 &  &  & \multicolumn{3}{c|}{Queries} & \multicolumn{2}{c|}{Relevant Docs/Query} & \multicolumn{2}{c}{Words/Doc}\\
Dataset &Corpus &Docs & Train & Valid & Test & Avg. & Std. & Avg. & Std.\\
\noalign{\vskip 1mm}
\hline
\noalign{\vskip 1mm}
%Wikipedia & Wikipedia & 5.9M & 1M & 20k & 20k \\
%aquaint-summa & aquaint & 1M & 700K & 20k & 20k \\
%aquaint2-summa & aquaint2 & 900k & 700K & 20k & 20k \\
TREC-CAR & Wikipedia Paragraphs & 3.5M & 585k & 195k & 195k & 3.6 & 5.7 & 84 & 68\\
Jeopardy & Wikipedia Articles & 5.9M & 118K & 10k & 10k & 1.0 & 0.0 & 462 & 990\\
MSA & Academic Papers & 480k & 270k & 20k & 20k & 17.9 & 21.5 & 165 & 158\\
%Robust 2004 & trec disks 4\&5 \& .gov & 530k & - & - & 250 \\
%Robust 2005 & aquaint  & 1M & - & - & 50 \\
%Web Track 2009-2012 & ClueWeb09-B & 50M & - & - & 200 \\
%\noalign{\vskip 1mm}
%\hline
\end{tabular}
\end{small}
\end{center}
\vskip -1mm
\caption{Summary of the datasets.}
\label{tab:datasets}

\vskip 4mm
\end{table*}

\subsection{Datasets}

We summarize in Table~\ref{tab:datasets} the datasets. 

\paragraph{TREC - Complex Answer Retrieval (TREC-CAR)} This is a publicly available dataset automatically created from Wikipedia whose goal is to encourage the development of methods that respond to more complex queries with longer answers~\cite{dietz2017trec}. A query is the concatenation of an article title and one of its section titles. The ground-truth documents are the paragraphs within that section. For example, a query is ``\textit{Sea Turtle, Diet}'' and the ground truth documents are the paragraphs in the section ``\textit{Diet}'' of the ``\textit{Sea Turtle}'' article. The corpus consists of all the English Wikipedia paragraphs, except the abstracts. The released dataset has five predefined folds, and we use the first three as the training set and the remaining two as validation and test sets, respectively.

\paragraph{Jeopardy} This is a publicly available Q\&A dataset introduced by~\citet{nogueira2016end}. A query is a question from the \textit{Jeopardy!} TV Show and the corresponding document is a Wikipedia article whose title is the answer. For example, a query is \textit{``For the last eight years of his life, Galileo was under house arrest for espousing this man’s theory''} and the answer is the Wikipedia article titled \textit{``Nicolaus Copernicus''}.
The corpus consists of all the articles in the English Wikipedia.

\paragraph{Microsoft Academic (MSA)} This dataset consists of academic papers crawled from Microsoft Academic API.\footnote{https://www.microsoft.com/cognitive-services/en-us/academic-knowledge-api} The crawler started at the paper~\citet{silver2016mastering} and traversed the graph of references until 500,000 papers were crawled. We then removed papers that had no reference within or whose abstract had less than 100 characters. We ended up with 480,000 papers.

A query is the title of a paper, and the ground-truth answer consists of the papers cited within. Each document in the corpus consists of its title and abstract.\footnote{This was done to avoid a large computational overhead for indexing full papers.}
%We plan to release this dataset publicly upon acceptance.
%This dataset differs from the other two in that it uses different corpus and terminologies, and it has more ground-truth documents per query, thus favoring reformulation methods that produce more comprehensive queries.

\begin{table*}[t]
\begin{center}
\begin{small}
\begin{tabular}{lccc|ccc|ccc}
 & \multicolumn{3}{c|}{TREC-CAR} & \multicolumn{3}{c|}{Jeopardy} & \multicolumn{3}{c}{MSA}\\
Method & R@40 & P@10 & MAP@40 & R@40 & P@10 & MAP@40 & R@40 & P@10 & MAP@40 \\
\noalign{\vskip 1mm}
\hline
\noalign{\vskip 1mm}
Raw-Lucene & 43.6 & 7.24 & 19.6 & 23.4 & 1.47 & 7.40 & 12.9 & 7.24 & 3.36 \\
Raw-Google & - & - & - & 30.1 & 1.92 & 7.71 & - & - & -\\
\noalign{\vskip 1mm}
\hline
\noalign{\vskip 1mm}
PRF-TFIDF & 44.3 & 7.31 & 19.9 & 29.9 & 1.91 & 7.65 & 13.2 & 7.27 & 3.50\\
PRF-RM & 45.1 & 7.35 & 19.5 & 30.5 & 1.96 & 7.64 & 12.3 & 7.22 & 3.38\\
PRF-Emb & 44.5 & 7.32 & 19.0 & 30.1 & 1.92 & 7.74& 12.2 & 7.22 & 3.20\\
Vocab-Emb & 44.2 & 7.30 & 19.1 & 29.4 & 1.87 & 7.80 & 12.0 & 7.21 & 3.21\\
\noalign{\vskip 1mm}
\hline
\noalign{\vskip 1mm}
%Syn-Emb\\
%Syn-RL\\
%Syn-SL\\
%Syn-Oracle\\
%\noalign{\vskip 1mm}
%\hline
%\noalign{\vskip 1mm}
SL-FF & 44.1 & 7.29 & 19.7 & 30.8 & 1.95 & 7.70 & 13.2 & 7.28 & 3.88\\
SL-CNN & 45.3 & 7.35 & 19.8 & 31.1 & 1.98 & 7.79 & 14.0 & 7.42 & 3.99\\
SL-Oracle & 50.8 & 8.25 & 21.0 & 38.8 & 2.50 & 9.92 & 17.3 & 10.12 & 4.89\\
\noalign{\vskip 1mm}
\hline
\noalign{\vskip 1mm}
RL-FF & 44.1 & 7.29 & 20.0 & 31.0 & 1.98 & 7.81 & 13.9 & 7.33 & 3.81\\
RL-CNN & 47.3 & 7.45 & 20.3 & 33.4 & \textbf{2.14} & 8.02 & 14.9 & 7.63 & 4.30\\
RL-RNN & \textbf{47.9} & \textbf{7.52} & \textbf{20.6}  & \textbf{33.7} & 2.12 & \textbf{8.07} & \textbf{15.1} & \textbf{7.68} & \textbf{4.35}\\
RL-RNN-SEQ & 47.4 & 7.48 & 20.3  & 33.4 & 2.13 & 8.01 & 14.8 & 7.63 & 4.27\\
RL-Oracle & 55.9 & 9.06 & 23.0 & 42.4 & 2.74 & 10.3 & 24.6 & 12.83 & 6.33
%\noalign{\vskip 1mm}
%\hline
%\noalign{\vskip 1mm}
\end{tabular}
\end{small}
\end{center}
\vskip -2mm
\caption{Results on Test sets. We use R@40 as a reward to the RL-based models.}
\label{tab:results}

%\vskip 4mm
\end{table*}

\subsection{Metrics and Reward}

Three metrics are used to evaluate performance:

\paragraph{Recall@K:} Recall of the top-K retrieved documents: 
\begin{equation}
	\text{R}@K = \frac{|D_K \cap D^*|}{|D^*|},
\end{equation}
where $D_K$ are the top-$K$ retrieved documents and $D^*$ are the relevant documents. Since one of the goals of query reformulation is to increase the proportion of relevant documents returned, recall is our main metric.

\paragraph{Precision@K:} Precision of the top-K retrieved documents:
\begin{equation}
	\text{P}@K = \frac{|D_K \cap D^*|}{|D_K|}
\end{equation}
Precision captures the proportion of relevant documents among the returned ones. Despite not being the main goal of a reformulation method, improvements in precision are also expected with a good query reformulation method. Therefore, we include this metric.

\paragraph{Mean Average Precision:} The average precision of the top-K retrieved documents is defined as:
\begin{equation}
	\text{AP}@K = \frac{\sum_{k=1}^K \text{P}@k \times \text{rel}(k)}{|D^*|},
\end{equation}
where
\begin{equation}
	\text{rel}(k) = 
    \begin{cases}
        1, & \text{if the k-th document is relevant;}\\
        0, & \text{otherwise.}
    \end{cases}
\end{equation}
The mean average precision of a set of queries $Q$ is then:
\begin{equation}
	\text{MAP}@K = \frac{1}{|Q|}\sum_{q \in Q} \text{AP}@K_q,
\end{equation}
where $\text{AP}@K_q$ is the average precision at $K$ for a query $q$. This metric values the position of a relevant document in a returned list and is, therefore, complementary to precision and recall. 

%We did not use metrics that taken into account the position of a document in a returned list, such as mean average precision (MAP) and discount cumulative gain (DCG), as the document order is not a main concern for the tasks evaluated in this work.\todo{not true for jeopardy and trec-car. true only if there will be other raking module in the pipeline.}

\paragraph{Reward}
We use $\text{R}@K$ as a reward when training the proposed RL-based models as this metric has shown to be effective in improving the other metrics as well.

\paragraph{SL-Oracle}
In addition to the baseline methods and proposed reinforcement learning approach, we report two oracle performance bounds. The first oracle is a supervised learning oracle (SL-Oracle). It is a classifier that perfectly selects terms that will increase performance according to the procedure described in Section~\ref{sec:proposed_methods}. This measure serves as an upper-bound for the supervised methods. Notice that this heuristic assumes that each term contributes independently from all the other terms to the retrieval performance. There may be, however, other ways to explore the dependency of terms that would lead to a higher performance.

\paragraph{RL-Oracle}
Second, we introduce a reinforcement learning oracle (RL-Oracle) which estimates a conservative upper-bound performance for the RL models. Unlike the SL-Oracle, it does not assume that each term contributes independently to the retrieval performance. It works as follows: first, the \textit{validation} or \textit{test} set is divided into $N$ small subsets $\{A_i\}_{i=1}^N$ (each with 100 examples, for instance).
An RL model is trained on each subset $A_i$ until it overfits, that is, until the reward $R_i^*$ stops increasing or an early stop mechanism ends training.\footnote{The subset should be small enough, or the model should be large enough so it can overfit.} Finally, we compute the oracle performance $R^*$ as the average reward over all the subsets: $R^*= \frac{1}{N}\sum_{i=1}^{N} R_i^*$.

This upper bound by the RL-Oracle is, however, conservative since there might exist better reformulation strategies that the RL model was not able to discover.

%The rationale for this choice comes from how search engines work: once a query is issued, a large set of documents is retrieved by matching words in an inverted index. The aim of this first step is to have a high recall. Then, a ranking function sorts these documents using more complex, slower, techniques. Since only a fraction of the sorted documents is returned, the goal of the ranking step is to have a high precision.

%Reformulating a query has the potential to retrieve relevant documents that otherwise would be missed if the proper terms were not used. Query reformulation can thus improve recall. It can also, however, improve precision. In fact, we show in the results section that RL-based models trained with recall as a reward also improve precision.\todo{rewrite the whole paragraph}

\begin{table}
\begin{center}
%\begin{small}
\begin{tabular}{lccc}
& TREC-CAR & Jeopardy & MSA \\
\noalign{\vskip 1mm}
\hline
\noalign{\vskip 1mm}
SL-Oracle & 13\% & 5\% & 11\% \\
RL-Oracle & 29\% & 27\% & 31\% \\
\end{tabular}
%\end{small}
\end{center}
\vskip -2mm
\caption{Percentage of relevant terms over all the candidate terms according to SL- and RL-Oracle.}

\label{tab:oracleterms}
%\vskip -2mm
\end{table}

\subsection{Implementation Details}

\paragraph{Search engine} We use Lucene and BM25 as the search engine and the ranking function, respectively, for all PRF, SL and RL methods. For Raw-Google, we restrict the search to the \textit{wikipedia.org} domain when evaluating its performance on the Jeopardy dataset. We could not apply the same restriction to the two other datasets as Google does not index Wikipedia paragraphs, and as it is not trivial to match papers from MS Academic to the ones returned by Google Search.

\begin{figure}
%\vskip 0.2in
\begin{center}
\centerline{\includegraphics[width=\columnwidth]{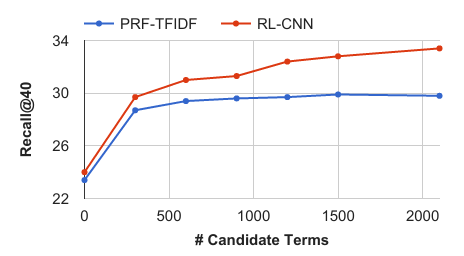}}
%\vskip -4mm
\caption{Our RL-based model continues to improve recall as more candidate terms are added, whereas a classical PRF method saturates.
}
\label{fig:feedback_terms}
\vskip -4mm
\end{center}

\end{figure}

%\paragraph{Source of candidate terms}
\paragraph{Candidate terms}
We use Wikipedia articles as a source for candidate terms since it is a well curated, clean corpus, with diverse topics.
%This may be due to the fact that the former is less noisy, as observed by~\citet{xu2009query}. Therefore, in our experiments a search engine returns two different sets of documents: one to compute the retrieval performance against the ground truth documents, and another for candidate terms.

%\paragraph{Sampling candidate terms}
At training and test times of SL methods, and at test time of RL methods, the candidate terms are from the first $M$ words of the top-$K$ Wikipedia articles retrieved. We select $M$ and $K$ using grid search on the validation set over $\{50,100,200,300\}$ and $\{1,3,5,7\}$, respectively. The best values are $M=300$ and $K=7$. These correspond to the maximum number of terms we could fit in a single GPU.

At training time of an RL model, we use only \textit{one} document uniformly sampled from the top-$K$ retrieved ones as a source for candidate terms, as this leads to a faster learning.

For the PRF methods, the top-$M$ terms according to a relevance metric (i.e., TF-IDF for PRF-TFIDF, cosine similarity for PRF-Emb, and conditional probability for PRF-RM) from each of the top-$K$ retrieved documents are added to the original query. We select $M$ and $K$ using grid search over $\{10, 50, 100, 200, 300, 500\}$ and $\{1, 3, 5, 9, 11\}$, respectively. The best values are $M=300$ and $K=9$.

\paragraph{Multiple Reformulation Rounds} Although our framework supports multiple rounds of search and reformulation, we did not find any significant improvement in reformulating a query more than once.
%, similarly to previous findings in the relevance feedback setting~\cite{harman1992relevance}
Therefore, the numbers reported in the results section were all obtained from models running two rounds of search and reformulation.

\paragraph{Neural Network Setup} 
For SL-CNN and RL-CNN variants, we use a 2-layer convolutional network for the original query. Each layer has a window size of 3 and 256 filters. We use a 2-layer convolutional network for candidate terms with window sizes of 9 and 3, respectively, and 256 filters in each layer. We set the dimension $d$ of the weight matrices $W,S,U$, and $V$ to $256$. For the optimizer, we use ADAM~\cite{kingma2014adam} with $\alpha=10^{-4}$, $\beta_1=0.9$, $\beta_2=0.999$, and $\epsilon=10^{-8}$. We set the entropy regularization coefficient $\lambda$ to $10^{-3}$.

For RL-RNN and RL-RNN-SEQ, we use a 2-layer bidirectional LSTM with 256 hidden units in each layer. We clip the gradients to unit norm. For RL-RNN-SEQ, we set the maximum possible number of generated terms to 50 and we use beam search of size four at test time.

We fix the dictionary of pre-trained word embeddings during training, except the vector for out-of-vocabulary words. We found that this led to faster convergence and observed no difference in the overall performance when compared to learning embeddings during training.

\section{Results and Discussion}

Table~\ref{tab:results} shows the main result. As expected, reformulation based methods work better than using the original query alone. Supervised methods (SL-FF and SL-CNN) have in general a better performance than unsupervised ones (PRF-TFIDF, PRF-RM, PRF-Emb, and Emb-Vocab), but perform worse than RL-based models (RL-FF, RL-CNN, RL-RNN, and RL-RNN-SEQ).

%Begin: added after the submission on apr/17
RL-RNN-SEQ performs slightly worse than RL-RNN but produces queries that are three times shorter, on average (15 vs 47 words). Thus, RL-RNN-SEQ is faster in retrieving documents and therefore might be a better candidate for a production implementation.
%End: added after the submission on apr/17

The performance gap between the oracle and best performing method (Table~\ref{tab:results}, RL-Oracle vs. RL-RNN) suggests that there is a large room for improvement. The cause for this gap is unknown but we suspect, for instance, an inherent difficulty in learning a good selection strategy and the partial observability from using a black box search engine.

\subsection {Relevant Terms per Document}

The proportion of relevant terms selected by the SL- and RL-Oracles over the total number of candidate terms (Table~\ref{tab:oracleterms}) indicates that only a small subset of terms are useful for the reformulation. Thus, we may conclude that the proposed method was able to learn an efficient term selection strategy in an environment where relevant terms are infrequent.

\subsection{Scalability: Number of Terms vs Recall}

Fig.~\ref{fig:feedback_terms} shows the improvement in recall as more candidate terms are provided to a reformulation method. The RL-based model benefits from more candidate terms, whereas the classical PRF method quickly saturates. In our experiments, the best performing RL-based model uses the maximum number of candidate terms that we could fit on a single GPU. We, therefore, expect further improvements with more computational resources.

\begin{figure*}
\begin{center}
%\vskip 0.2in
\centerline{\includegraphics[width=0.68\paperwidth]{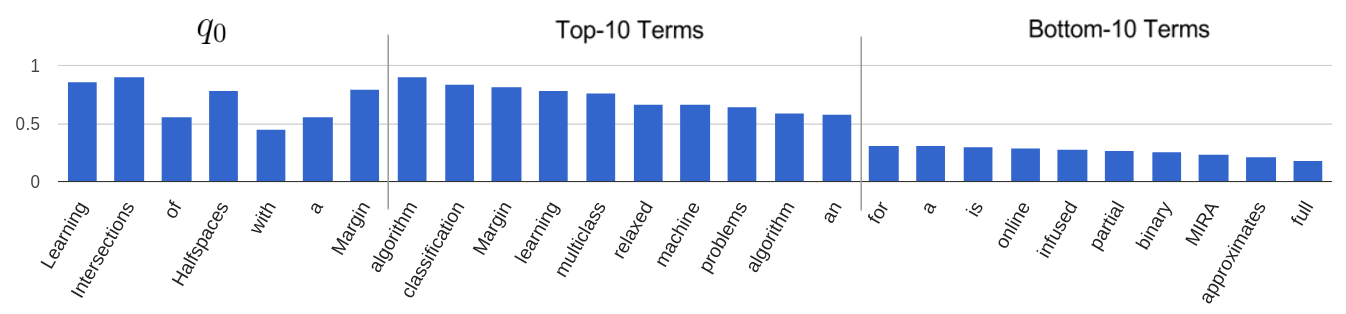}}
\vskip 0.1in
\centerline{\includegraphics[width=0.73\paperwidth]{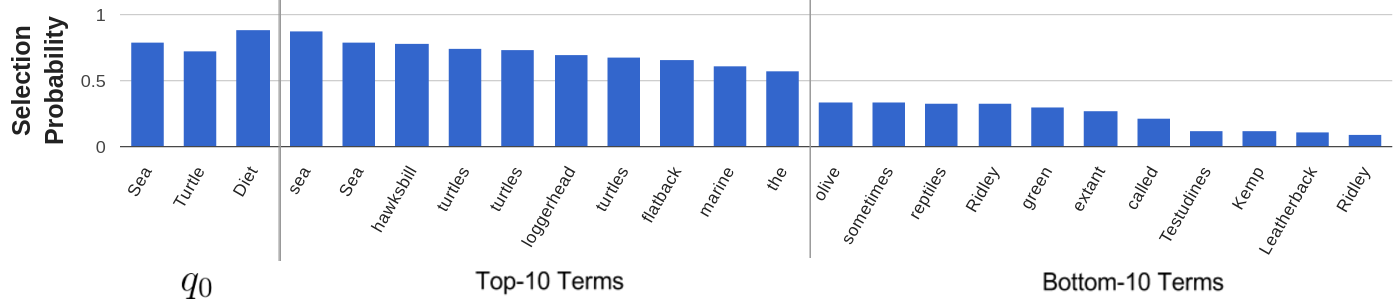}}
\vskip -0.1in
\caption{Probabilities assigned by the RL-CNN to candidate terms of two sample queries: ``\textit{Learning Intersections of Halfspaces with a Margin}'' (top) and ``\textit{Sea Turtle Diet}'' (bottom). We show the original query terms and the top-10 and bottom-10 document terms with respect to their probabilities. %We use in a reformulated query all the terms with a probability higher than 0.5.
% Thus, the reformulated queries are ``\textit{Sea Turtle Diet Sea turtles marine turtles are the seven species sea turtles the loggerhead ridley hawksbill flatback}'' (top) and ``\textit{be a house the Swiss Alps, a ski lodge built in ski hill, chalet style Swiss chalet.}'' (bottom). 
}
\label{fig:probs}
\end{center}

\vskip -4mm
\end{figure*}

\begin{table}
%\vskip -4mm
\begin{center}
\begin{small}
\begin{tabular}{l|l}
Query & Top-3 Retrieved Documents \\
\noalign{\vskip 1mm}
\hline
\noalign{\vskip 1mm}
(Original) \textit{The Cross} & \textit{-The Cross Entropy Method} \\
\textit{Entropy Method for} & \textit{for Network Reliability Estim.}\\
\textit{Fast Policy Search} & \textit{\textbf{-Robot Weightlifting by}}\\
& \textbf{\textit{Direct Policy Search}}\\
& \textit{-Off-policy Policy Search}\\
\noalign{\vskip 1mm}
\hline
\noalign{\vskip 1mm}
(Reformulated) \textit{Cross } & \textbf{\textit{-Near Optimal Reinforcement}}\\
\textit{Entropy Fast Policy } & \textbf{\textit{Learning in Polynom. Time}}\\
\textit{Reinforcement } & \textit{-The Cross Entropy Method}\\
\textit{Learning policies} & \textit{for Network Reliability Estim.}\\
\textit{global search}& \textit{\textbf{-Robot Weightlifting by}}\\
\textit{optimization biased} & \textbf{\textit{Direct Policy Search}}\\
\noalign{\vskip 2mm}
\hline
\noalign{\vskip 1mm}
\hline
\noalign{\vskip 1mm}
(Original) \textit{Daikon} & ``\textit{...many types of pickles are}\\
\textit{Cultivation} & \textit{made with daikon, includ...}''\\
& ``\textbf{\textit{Certain varieties of daikon}}\\
& \textbf{\textit{can be grown as a winter...}}''\\
& ``\textit{In Chinese cuisine, turnip}\\
& \textit{cake and chai tow kway...}''\\
\noalign{\vskip 1mm}
\hline
\noalign{\vskip 1mm}
(Reformulated) \textit{Daikon} & ``\textit{...many types of pickles are}\\
\textit{Cultivation root seed } & ``\textit{made with daikon, includ...}''\\
\textit{grow fast-growing} & ``\textbf{\textit{Certain varieties of daikon}}\\
\textit{Chinese leaves} & \textbf{\textit{can be grown as a winter...}}''\\
& ``\textbf{\textit{The Chinese and Indian }}\\
& \textbf{\textit{varieties tolerate higher....}}''
\end{tabular}
\end{small}
\end{center}
%\vskip 2mm
\caption{Top-3 retrieved documents using the original query and a query reformulated by our RL-CNN model. In the first example, we only show the titles of the retrieved MSA papers. In the second example, we only show some words of the retrieved TREC-CAR paragraphs. \textbf{Bold} corresponds to ground-truth documents.}
\label{tab:return_example}

%\vskip -2mm
\end{table}

\begin{table}
\begin{center}
\begin{small}
\begin{tabular}{l|l}
Trained on & Selected Terms\\
\noalign{\vskip 1mm}
\hline
\noalign{\vskip 1mm}
TREC-CAR & \textit{serves american national Winsted}\\
& \textit{accreditation}\\
\noalign{\vskip 1mm}
\hline
\noalign{\vskip 1mm}
Jeopardy & \textit{Tunxis Quinebaug Winsted NCCC}\\
\noalign{\vskip 1mm}
\hline
\noalign{\vskip 1mm}
MSA & \textit{hospital library arts center cancer center}\\
& \textit{summer programs}
\end{tabular}
\end{small}
\end{center}
%\vskip -0.6mm
\caption{Given the query \textit{``Northwestern Connecticut Community College''}, models trained on different tasks choose different terms.}
\label{tab:samequery}
%\vskip -0.4mm
\end{table}

%\subsection{Synonyms as Candidate Terms}
%\label{sec:syns}
%A dictionary of synonyms, such as Wordnet~\cite{miller1995wordnet}, can also be a source of candidate terms. To test if this could improve performance, we did the following experiment: for each word in the original query we added to the pool of candidate terms the 20 most  similar words according to WordNet. We then measured both SL- and RL-Oracle performances and found almost no improvement over using the original query alone (Raw-Lucene). The same outcome was observed in all three datasets.

%From this result, together with similar previous findings in the literature~\cite{xu1996query}, we decided to not further experiment with this type of source for candidate terms.\todo{not necessary?}

%This result indicates that either the synonyms are an inadequate source of terms for the datasets/tasks used in this work or our current models cannot learn to use them efficiently, or both. However, more study is needed before any further conclusion.

\subsection {Qualitative Analysis}

We show two examples of queries and the probabilities of each candidate term of being selected by the RL-CNN model in Fig.~\ref{fig:probs}.

Notice that terms that are more related to the query have higher probabilities, although common words such as "\textit{the}" are also selected. This is a consequence of our choice of a reward that does not penalize the selection of neutral terms.

%Also notice in the second example (bottom) that the selected word ``\textit{chalet}'' is the answer to the Jeopardy question. This leads us to the insight that the best reformulation of a question is its \textit{answer}.

In Table~\ref{tab:return_example} we show an original and reformulated query examples extracted from the MS Academic and TREC-CAR datasets, and their top-3 retrieved documents. Notice that the reformulated query retrieves more relevant documents than the original one. As we conjectured earlier, we see that a search engine tends to return a document simply with the largest overlap in the text, necessitating the reformulation of a query to retrieve semantically relevant documents.

\paragraph{Same query, different tasks} We compare in Table~\ref{tab:samequery} the reformulation of a sample query made by models trained on different datasets. The model trained on TREC-CAR selects terms that are similar to the ones in the original query, such as ``\textit{serves}'' and ``\textit{accreditation}''. These selections are expected for this task since similar terms can be effective in retrieving similar paragraphs. On the other hand, the model trained on Jeopardy prefers to select proper nouns, such as ``\textit{Tunxis}'', as these have a higher chance of being an answer to the question. The model trained on MSA selects terms that cover different aspects of the entity being queried, such as ``\textit{arts center}'' and ``\textit{library}'', since retrieving a diverse set of documents is necessary for the task the of citation recommendation.

\subsection {Training and Inference Times}
Our best model, RL-RNN, takes 8-10 days to train on a single K80 GPU. At inference time, it takes approximately one second to reformulate a batch of 64 queries. Approximately 40\% of this time is to retrieve documents from the search engine.

\section{Conclusion}

We introduced a reinforcement learning framework for task-oriented automatic query reformulation. An appealing aspect of this framework is that an agent can be trained to use a search engine for a specific task. The empirical evaluation has confirmed that the proposed approach outperforms strong baselines in the three separate tasks. The analysis based on two oracle approaches has revealed that there is a meaningful room for further development. In the future, more research is necessary in the directions of (1) iterative reformulation under the proposed framework, (2) using information from modalities other than text, and (3) better reinforcement learning algorithms for a partially-observable environment.

%\subsection{Future Work}

%The performance gap between the SL and RL Oracles indicates that terms should not be treated as contributing independently to the retrieval performance. Therefore, we suggest that future works should focus on methods that take into account the interaction between terms to improve retrieval efficacy.

%A future improvement is that of reducing querying time, as it represents 90\% of training time. Thus, making retrieval faster would allow more gradient updates for the same amount of training, which can, in turn, lead to an increase the retrieval performance.\todo{remove?}

%Another future direction is the use of sequence-to-sequence models~\cite{sutskever2014sequence,bahdanau2014neural} to produce a reformulated query. These models can output terms other than the ones from the retrieved documents and this may lead to better reformulations.

\section*{Acknowledgements} 

RN is funded by Coordenação de Aperfeiçoamento de Pessoal de Nível Superior (CAPES). KC thanks support by Facebook, Google and NVIDIA. This work was partly funded by the Defense Advanced Research Projects Agency (DARPA) D3M program. Any opinions, findings, and conclusions or recommendations expressed in this material are those of the authors and do not necessarily reflect the views of DARPA. 

\bibliography{main}
\bibliographystyle{emnlp_natbib}

\end{document}